\documentclass[useAMS,usenatbib]{mn2e}

\usepackage{epsfig}
\usepackage{aas_macros}

\def\Msun{\ifmmode{~M_\odot}\else$M_\odot$~\fi}
\def\kms{\ifmmode{\,\mbox{km\,s$^{-1}$}}\else\,km\,s$^{-1}$\fi}
\def\kpc{\ifmmode{\,\mbox{kpc}}\else\,kpc\fi}

\def\Msun{\ifmmode{~M_\odot}\else$M_\odot$~\fi}
\def\Mden{\ifmmode{~M_\odot $pc$^{-3}}\else$M_\odot $pc$^{-3}$\fi}

\def\Msun{\ifmmode{~{\rm M}_\odot}\else${\rm M}_\odot$~\fi}
\def\etal{\ifmmode{${et~al.}$}\else {et~al.}\fi}
\def\Mden{\ifmmode{~{\rm M}_\odot~{\rm pc}^{-3}}
 \else${\rm M}_\odot$~pc$^{-3}$\fi}

\def\hc{{H$\gamma$}~}
\def\hd{{H$\delta$}~}
\def\siglos{\ifmmode{\sigma_{\rm LOS}}\else$\sigma_{\rm LOS}$\fi}
\def\sigsq{\ifmmode{{\bmath \sigma}^2}\else${\bmath \sigma}^2$\fi}
\def\sig{\ifmmode{{\bmath \sigma}}\else${\bmath \sigma}$\fi}
\newcommand{\degree}{\mbox{$^{\circ}$}}
\def\num{530}
\def\chisq{$\chi^2$}
\title[Kinematics of the Galactic Halo]{Kinematics of the Galactic
Halo from Horizontal Branch stars in the Hamburg/ESO Survey}

\author[C.Thom, et~al.]{Christopher Thom$^{1,2}$,  Chris Flynn$^{2}$, Michael S. Bessell$^{3}$, Jyrki H{\"a}nninen$^{2}$, 
\newauthor Timothy C. Beers$^{4}$,  Norbert Christlieb$^{5}$, Dionne James$^{6}$, Johan Holmberg$^{2}$,
\newauthor Brad K. Gibson$^{1}$ \\
$^{1}$ Astrophysics and Supercomputing Centre, Swinburne University, Melbourne,
Australia\\
$^{2}$ Tuorla Observatory, V\"ais\"al\"antie 20, FI-21500, Piikki\"o, Finland\\
$^{3}$ Mt Stromlo and Siding Spring Observatory, ANU, Canberra, Australia \\
$^{4}$ Department of Physics \& Astronomy, and JINA: Joint Institute for Nuclear
Astrophysics, \\ 
\hspace{2ex}Michigan State University, E. Lansing, MI, 48824, USA \\
$^{5}$ Hamburger Sternwarte, Gojenbergsweg 112, D-21029, Hamburg, Germany \\
$^{6}$ Anglo-Australian Observatory, PO Box 296, Epping, NSW 1710, Australia \\
}

\begin{document}

\date{Accepted  Received ; in original form }

\pagerange{\pageref{firstpage}--\pageref{lastpage}} \pubyear{2004}

\maketitle

\label{firstpage}

\begin{abstract}
  Large samples of Field Horizontal Branch (FHB) stars make excellent
  tracers of the Galactic halo; by studying their kinematics, one can infer
  important physical properties of our Galaxy.  Here we present the results
  of a medium-resolution spectroscopic survey of \num~FHB stars selected
  from the Hamburg/ESO survey. The stars have a mean distance of $\sim
  7\kpc$ and thus probe the inner parts of the Milky Way halo. We measure
  radial velocities from the spectra in order to test the model of
  Sommer-Larsen et al., who suggested that the velocity ellipsoid of the
  halo changes from radially-dominated orbits to tangentially-dominated
  orbits as one proceeds from the inner to the outer halo.  We find that
  the present data are unable to discriminate between this model and a more
  simple isothermal ellipsoid; we suggest that additional observations
  towards the Galactic centre might help to differentiate them.
\end{abstract}

\begin{keywords}
stars: horizontal branch -- Galaxy: halo -- Galaxy: kinematics and dynamics
\end{keywords}

\section{Introduction}
\label{sec: intro}

The stellar halo only comprises about 1\,\% of the luminous mass of the
Galaxy and is composed of very old, metal-poor stars. In spite of this, the
present-day dynamical and chemical state of the stellar halo has exerted
considerable influence in shaping our understanding of the formation of
large disk galaxies such as the Milky Way, particularly in its very early
stages.

Large samples of Field Horizontal Branch stars (FHBs) in the Galactic halo
are excellent ``test particles'' for studies of halo kinematics
\citep[e.g.][]{slfc94,sbfwc97,sirko04b}; tracing the mass
\citep{beers04-mwmass} and the merger history \citep{ksk94,brown04-2mass}
of the Milky Way. Their intrinsic brightness and the relative ease with
which distances can be obtained, together with their relatively clean
spectra, also make them ideal probes of distances to High Velocity Clouds
(e.g. \citealt{schwarz95}; Thom \etal, in preparation).

The HK survey of Beers and collaborators \citep[e.g.][]{beers85-HK,
beers92-HK} has identified $\sim 12,000$ candidate Horizontal Branch stars,
of which about half are expected to be on the Horizontal Branch; the other
half appear to be a mixture of A-type main-sequence stars and halo blue
stragglers \citep{beers88-HKcat,beers96-HKcat}.  More recently, several
groups have reported large and clean samples of Horizontal Branch stars.
The Hamburg/ESO survey (HES), originally designed to identify
low-luminosity quasars, has been shown to contain a remarkable sample of
interesting stars, including 8321 FHBs. This sample has a contamination
level by non-FHBs (mostly high-gravity A-type stars) of less than 16\,\%
\citep{nc04-fhb}. A subset of these stars is used here to constrain models
of halo kinematics.

The Sloan Digital Sky Survey (SDSS), although primarily designed for
extragalactic studies, has also revealed a large sample of 1170 Horizontal
Branch stars at distances up to $\sim 100\kpc$ \citep{sirko04a}. From this
sample \citet{sirko04b} measured an isotropic velocity ellipsoid for the
outer halo. This result contrasts with halo stars in the solar
neighbourhood, which show an ellipsoid elongated in the radial direction
\citep{chiba&beers00, gould03,gould04-erratum03}.

\citet{wilhelm99a} have described a technique for separating Horizontal
Branch stars from higher gravity A stars on the basis of broadband {\it
UBV} colours and medium-resolution spectroscopy. \citet{clewley02,
clewley04} explored similar techniques based on broadband colours and
medium-resolution spectroscopy, and also presented a method relying solely
on medium resolution spectroscopy. They used this method to identify $\sim
100$ stars at distances of $>30\kpc$ with the aim of providing better
constraints on the mass of the Milky Way.

\begin{figure}
  \begin{center}
    \leavevmode
    \epsfig{file=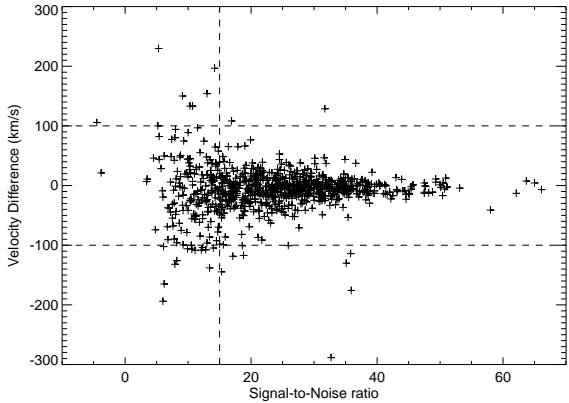,width=8cm}
    \caption{Radial velocity differences, as measured using the \hd and \hc
    lines, versus Signal-to-Noise ratio of the spectra, for our sample
    stars. The vertical and horizontal lines indicate the adopted $S/N >
    15$ and $|v_{\mbox{\scriptsize diff}}| < 100\kms$ limits, respectively (see text).}
    \label{fig: diff-vel-snr}
  \end{center}
\end{figure}

\citet{brown04-2mass} used the full {\it 2MASS} point-source catalogue to
select close to 100,000 objects, of which they expect 47\,\% to be on the
Blue Horizontal Branch. They find an absence of structure in the inner halo
(the sample has distance $d < 9\kpc$), concluding that there have been no
major accretion events in the inner halo over the past few Gyrs.

Here we present the results of a programme to obtain radial velocities
(RVs), and study the kinematics, of more than 500 FHB stars selected from
the HES catalogue.  We wish to test specifically the predictions of the
\citet{slfc94} (hereafter, SLFC) model, as refined by \citet{sbfwc97}.  In
the SLFC model, the velocity ellipsoid of the halo changes from
radially-dominated orbits to tangentially-dominated orbits as one proceeds
from the inner to the outer halo.  We attempt to distinguish between this
model and the simpler isothermal halo model, which specifies an isotropic
distribution of stellar orbits, independent of Galactocentric radius, in
the outer halo.  Section 2 presents the sample and observations. It also
includes details on the radial velocity measurements and selection
criteria.  Section 3 discusses the models which we wish to test. Section 4
details the final sample selection, division into fields, and calculation
of the line-of-sight velocity dispersions, \siglos. Section 5 presents the
major results and analysis. A summary and conclusions are provided in
Section 6.

\begin{figure}
  \begin{center}
    \leavevmode
    \epsfig{file=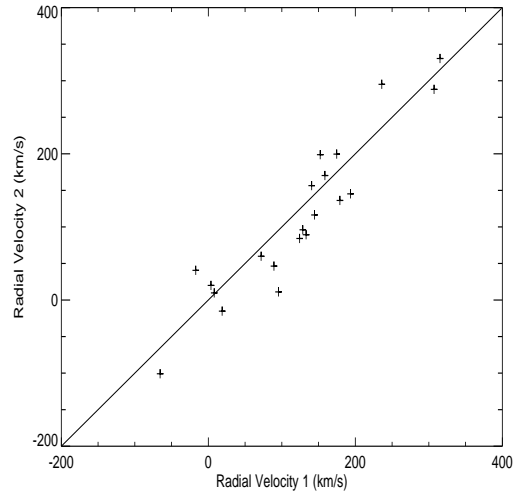,width=70mm,height=70mm}
    \caption{RV1 versus RV2 for the 21 stars with repeated
    measurements. The solid line is the one-to-one correspondence.}
    \label{fig: doubles}
  \end{center}
\end{figure}

%%%%%%%%%%%%%%%%%%%%%%%%%%%%%%%%%%%%%%%%%%%%%%%%%%%%%%%%%%%%%%%%%%%%%%%

\section{Sample and Observations}
\label{sec: observations}

Our sample of stars was drawn from the \citet{nc04-fhb} catalogue of FHB
stars selected from the Hamburg/ESO survey. As part of a large programme of
follow-up observations of the stellar content of the HES and HK surveys,
over 1100 FHB stars were observed at the UK Schmidt Telescope
(UKST)\footnote{The UKST is operated and supported by the Anglo-Australian
Observatory, on behalf of the astronomical communities of Australia and the
UK}, using the six degree field (6dF) fibre-fed spectrograph. The 6dF
instrument allows one to observe up to 150 objects simultaneously across
the 6$\degree$ field of the Schmidt telescope, obtaining spectra at
2\,{\AA} resolution. More than forty nights of observations took place
between April 2001 and April 2003.  Integration times were typically
$4$--$5$ times $2700$\,s per field, depending on observing conditions.

The targets were, in most cases, selected by HES field number, with roughly
equal numbers of fibres used for metal-poor candidates and FHB stars.  The
data were reduced using the {\it Figaro} and {\it 6dfdr} software packages,
the latter being written specifically to handle the fibre data from the 6dF
instrument.  This process yielded {\it FITS} files containing the reduced
spectra of all 150 fibres.  These files were separated, with one spectrum
per file and assigned a unique name based on the star name, field number,
and date of observation. A hand-screening process was then used to exclude
obvious problems and artefacts. 

Catalogues of FHB stars may contain a significant fraction of contaminants
-- higher surface gravity A-type stars which fall within the colour range
of the survey (in the case of the HES, $-0.2 < B-V < 0.3$).  Since the HES
FHB candidates have been shown to have a much lower level of contamination
from higher-gravity stars \citep[$< 16\,\%$;][]{nc04-fhb}, as compared to
the HK survey \citep[$\sim 50\,\%$;][]{beers96-HKcat}, we considered only
the HES stars in our observations.  We made no attempt to separate the
potential contaminants from the true FHB stars, judging that a $\sim
16\,\%$ level is acceptable for our present purpose. While it is, in
principle, possible to further refine the classification, using the
spectroscopic method of \citet{clewley02} would result in a sample clean to
about the $\sim 12\,\%$ level and hence nothing would be gained.  Since
systematic, accurate photometry is not available, methods based on this are
not applicable.  Conversely, we have no way of obtaining RVs from the
objective prism of the HES, but can measure velocities in the 6df spectra,
since the S/N requirements for RVs are less stringent than those for
classification.

%%%%%%%%%%%%%%%%%%%%%%%%%%%%%%%%%%%%%%%%%%%%%%%%%%%%%%%%%%%%%%%%%%%%%%%%%5

\begin{figure}
  \begin{center}
    \leavevmode
    \epsfig{file=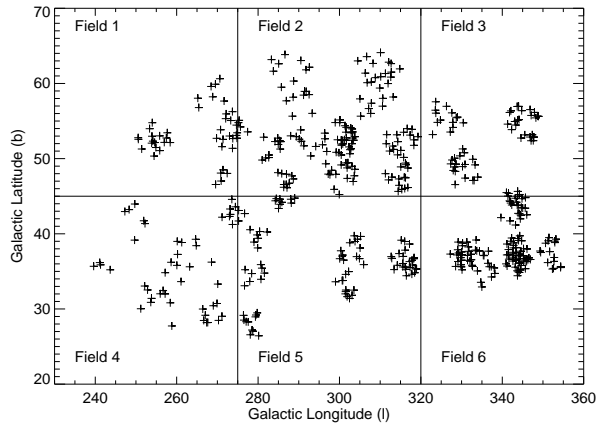,width=8cm}
    \caption{Sky distribution of all \num~stars in 6 northern fields.}
    \label{fig: allsky-lb}
  \end{center}
\end{figure}

\subsection{Radial Velocities and Selection}
\label{sec: rv}

Sersic profiles were fit to the \hc and \hd Balmer lines, as described by
\citet{clewley02}.  Due to changes in the spectrograph settings over the
long observing period, these were the only two prominent lines common to
all spectra. We rejected any star for which we could not obtain a radial
velocity measurement in both lines.  We also noticed large systematic
discrepancies ($>150\kms$) between the measurements of RV from the \hc and
\hd lines in some stars. These stars were noted to be concentrated in one
field (HES field 573), hence all stars in this field were rejected. One
star was also rejected due to difficulties with sky subtraction. This left
a total of 827 stars.

We sought to combine the velocity measurements from the two lines into an
average value, but we first used the difference in the measurements of the
two lines to provide an internal consistency check on the radial
velocities.  Fig.~\ref{fig: diff-vel-snr} shows a plot of the difference
between the \hd and \hc line, versus signal-to-noise ratio. We applied
cuts, retaining all stars with measurement difference less than 100\kms~and
S/N greater than 15, as shown in Fig~\ref{fig: diff-vel-snr}. An average RV
was then computed from all available measurements. Heliocentric corrections
were calculated using the IRAF task {\it rvcorrect}. This selection yielded
636 RV measurements.

Twenty-one stars in our sample were observed at more than a single epoch
and thus enable a further check on the measurements. The correlation
between the two average RVs for these stars is shown in Fig.~\ref{fig:
doubles}. The difference between these two measurements for a given star --
the range of measurements -- provides an estimate of the standard deviation
of the underlying error distribution \citep{pearson-26-range-outliers,
tippett-25-range-outliers}. This distribution will be normally distributed
with $\mu = RV$ and $\sigma = $ error in the measurement. Applying a
correction factor of 1.12838 (see Table X of
\citealt{tippett-25-range-outliers}) to the absolute value of the range, we
obtain a mean statistical velocity error of 30\kms. This agrees well with
the assumption of Gaussian errors, under which we obtain 27\kms, dividing
the standard deviation of the velocity differences by $\sqrt{2}$.  We
therefore adopt 30\kms\ as the formal error on our radial velocity
measurements. To definitively characterise the error on a single RV
measurement would require many repeated velocity measurements of the same
star. For the analysis, the two RV measurements were combined into a single
entry by averaging the two independent RVs, weighted by the S/N in the
respective spectra, yielding RVs for 615 unique stars.

\begin{figure}
  \begin{center}
    \leavevmode
    \epsfig{file=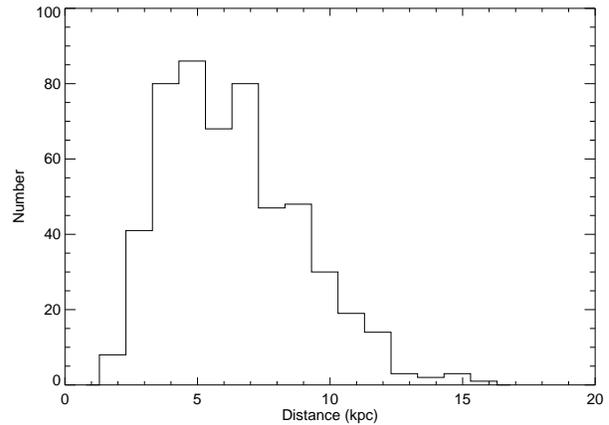,width=8cm}
    \caption{Histogram of heliocentric distances for the final sample of
\num~stars.}
    \label{fig: dist-hist}
  \end{center}
\end{figure}

%%%%%%%%%%%%%%%%%%%%%%%%%%%%%%%%%%%%%%%%%%%%%%%%%%%%%%%%%%%%%%%%%%%%%%%%%

\begin{figure*}
  \begin{center}
    \leavevmode
    \epsfig{file=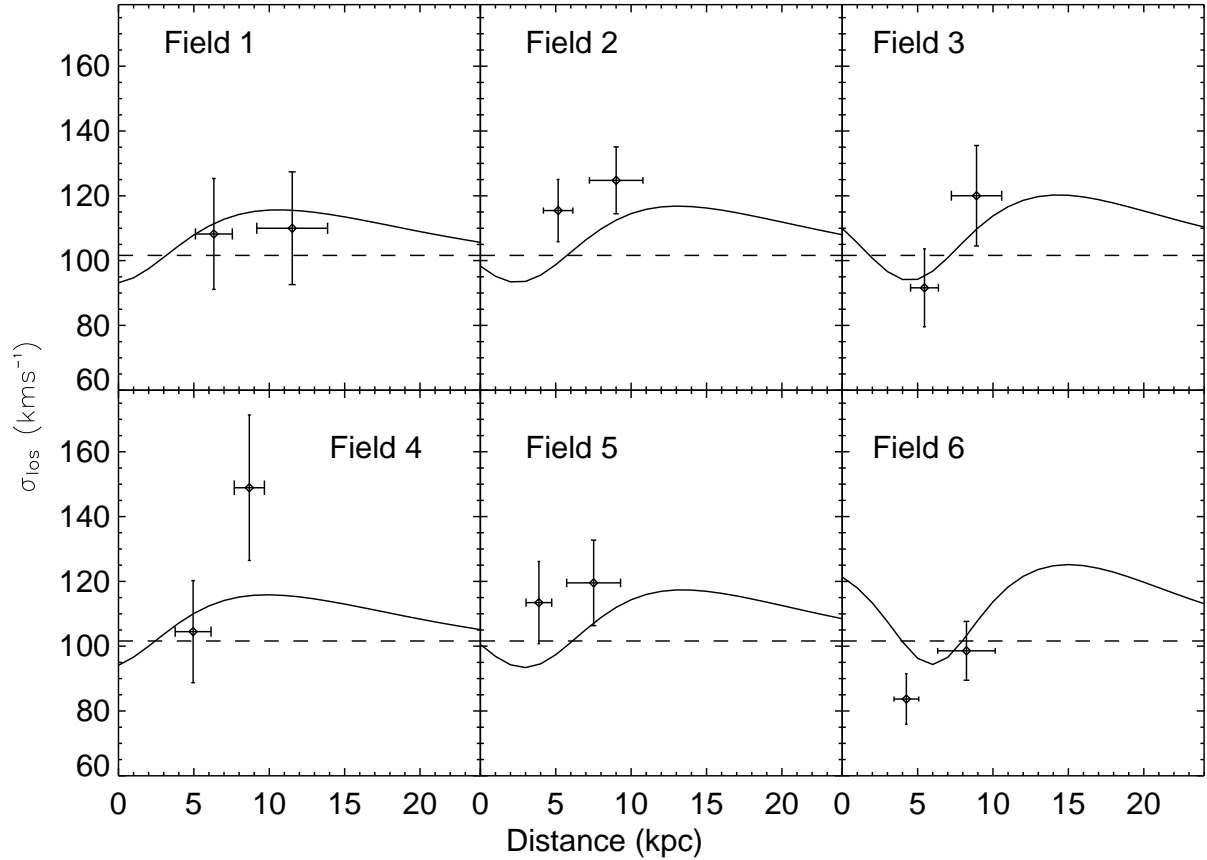,width=16cm}

    \caption{Comparison of measured \siglos\ to the two models. The
    horizontal dashed line shows an isothermal model with the value
    measured by Sirko \etal. The solid line is the line-of-sight velocity
    dispersion predicted by the SLFC model for the average line-of-sight of
    the field.  The error bars show the 1$\sigma$ errors.}

    \label{fig: sigma-plots}
  \end{center}
\end{figure*}

\section{Kinematic Models}
\label{models}

The density distribution of the stellar halo follows a power-law,
$R^{-\alpha}$, with Galactocentric radius, $R$. The value of $\alpha$ has
been the subject of many studies, using RR~Lyrae stars, halo giants, and
Blue Horizontal Branch (BHB) stars \citep[e.g.][]{preston-shectman-beers91,
ksk94, sluis98-halo-density, morrison00-spaghetti}. Most of these converge
on a value of $\alpha \approx 3.0\,$--$\,3.5$.  In contrast,
\citet{robin-etal-00-halo-flattening} find a lower value consistent with
the range $\alpha = 2.0\,$--$\,2.75$, whilst noting significant degeneracy
between $\alpha$ and the flattening parameter $c/a$.  This contrasts with
the density distribution of the Galactic dark halo, which follows $R^{-2}$
(i.e., the isothermal case expected from a flat rotation curve), at least
in the outer parts where it dominates over the luminous matter.

Stars in the halo of the Milky Way are drawn from a population with a total
dispersion on the order of 200\kms, i.e. they have velocities needed to
maintain equilibrium with the Galactic gravitational potential. In the
stellar halo, this velocity is directed roughly equally into three
components, resulting in a system with little bulk rotation and thus mainly
``pressure support'' against gravitational collapse.

The three components of the stellar halo's velocity ellipsoid have been
measured accurately in the solar neighbourhood, for stars within roughly
1\kpc~of the Sun, to be $\sigma = (\sigma_R, \sigma_\theta, \sigma_\phi)
\approx (140, 100, 100)\kms$ \citep{chiba&beers00,gould03}, where we have
adopted a spherical coordinate system $(R,\theta,\phi)$ around the Galactic
centre. This determination is based on stars for which all three components
of the space velocity can be measured via both radial line-of-sight
velocities and proper motions. Beyond a few kpc from the Sun, usually only
the radial velocity can be measured (a situation which will change
dramatically when the ESA GAIA mission is completed;
\citealt{perryman02-gaia}), and special techniques are required to
reconstruct the three components of the velocity ellipsoid. One can either
propose and fit data to models of the velocity ellipsoid as a function of
position in the Galaxy -- as performed by \citet{sbfwc97} for some $\sim
700$ stars -- or attempt to recover the three components directly
\citep{sirko04b}.

The velocity ellipsoid of the stellar halo is a consequence of its density
profile and the total matter distribution in the Galaxy. Let us assume that
the 3-D dispersion of the velocity distribution is isothermal (i.e., it has
the same magnitude in all three components), and that the density falls off
like $R^{-\alpha}$. It then follows, for a Galactic potential dominated by
a dark halo with a flat rotation curve (characterised by constant circular
velocity $V_0$) that $\sigma = {V_0}/\sqrt{\alpha}$. For the Milky Way,
$V_0 \approx 220$ \kms, and thus $\sigma \approx 120\kms$ for each
component \citep{binney&tremaine87}.  The total velocity dispersion is then
of order $\sqrt{3} \sigma \approx 200\kms$.

The stellar halo is clearly not isothermal in the vicinity of the Sun (the
component of the velocity dispersion in the radial direction is $\approx$
40\,\% larger than the two other components of the velocity ellipsoid);
nevertheless, the total velocity dispersion is of order 200\kms.

What is the velocity dispersion like elsewhere in the halo? SLFC and
\citet{sirko04b} have attempted a reconstruction from radial velocities of
distant halo stars. SLFC proposed a model for the velocity dispersion
$\sigma(R,\theta, \phi)$ as a function of Galactocentric radius $R$ alone;
the model permitted sharp deviations from isothermal behaviour (such as is
seen in the solar neighbourhood) and was fit to both the local and distant
halo stars. This model was subsequently refined using a much larger sample
of about $\sim 700$ distant halo stars by \citet{sbfwc97}. It was found
that the radial anisotropy ($\sigma_R > \sigma_{\phi,\theta}$) seen near
the Sun persists into the inner halo ($R < 8\kpc$)), changing to tangential
anisotropy in the outer halo (roughly $R>15-20\kpc$). \citet{sirko04b}
present an ostensibly opposing view; from radial velocities of 1170 halo
stars found in the SDSS, they found that the velocities of (distant) halo
stars are very close indeed to isothermal $(\sigma_R,\sigma_\theta,
\sigma_\phi) = (101.4 \pm 2.8, 97.7 \pm 16.4, 107.4 \pm 16.6)\kms$. The
\citeauthor{sirko04b} study is essentially of the outer halo, where the
SLFC model is not yet well constrained, hence the ostensible contradiction
with the non-isothermal model of SLFC is, as those authors argued, probably
not significant.

We treat the halo's kinematics statistically, characterised by the velocity
ellipsoid alone; i.e. we explicitly ignore substructure, such as dissolving
satellites or other features in density and/or velocity space. From the
point of view of halo substructure, the halo can be divided into inner and
outer regions. The inner halo is that part which is within $\sim 15\kpc$
radius from the Galactic centre. In this region, the relevant dynamical
time scales are expected to be considerably shorter than the age of the
Galaxy. This expectation is borne out by the lack of detected substructure
\citep[e.g.][]{brown04-2mass}.  Thus, in the inner halo, the stellar
population appears likely to be fairly well mixed, with a smooth density
distribution.  There is additional evidence that this is the case in the
solar neighbourhood, based on an analysis of 4588 sub-dwarfs
\citep{gould03,gould04-erratum03}. In the outer halo, beyond some 20\kpc~or
so, dynamical time scales approach the age of the Galaxy. In these regions
it has long been suspected that the halo would contain the debris of
dissolving satellite galaxies; this is now well established
(e.g. \citealt{majewski94,ibata94,lynden-bell95,morrison00-spaghetti}).
While the outer halo is undoubtedly quite lumpy, most of the stellar halo's
mass, and most of that part probed via our data, is in the well mixed inner
halo. Additional tests of this assertion will be the subject of a
forthcoming study.

%%%%%%%%%%%%%%%%%%%%%%%%%%%%%%%%%%%%%%%%%%%%%%%%%%%%%%%%%%%%%%%%%%%%%%%%%

\section{Final Sample and Velocity Dispersion}
\label{sec: fbs}

\begin{figure*}
  \begin{center}
    \leavevmode
    \epsfig{file=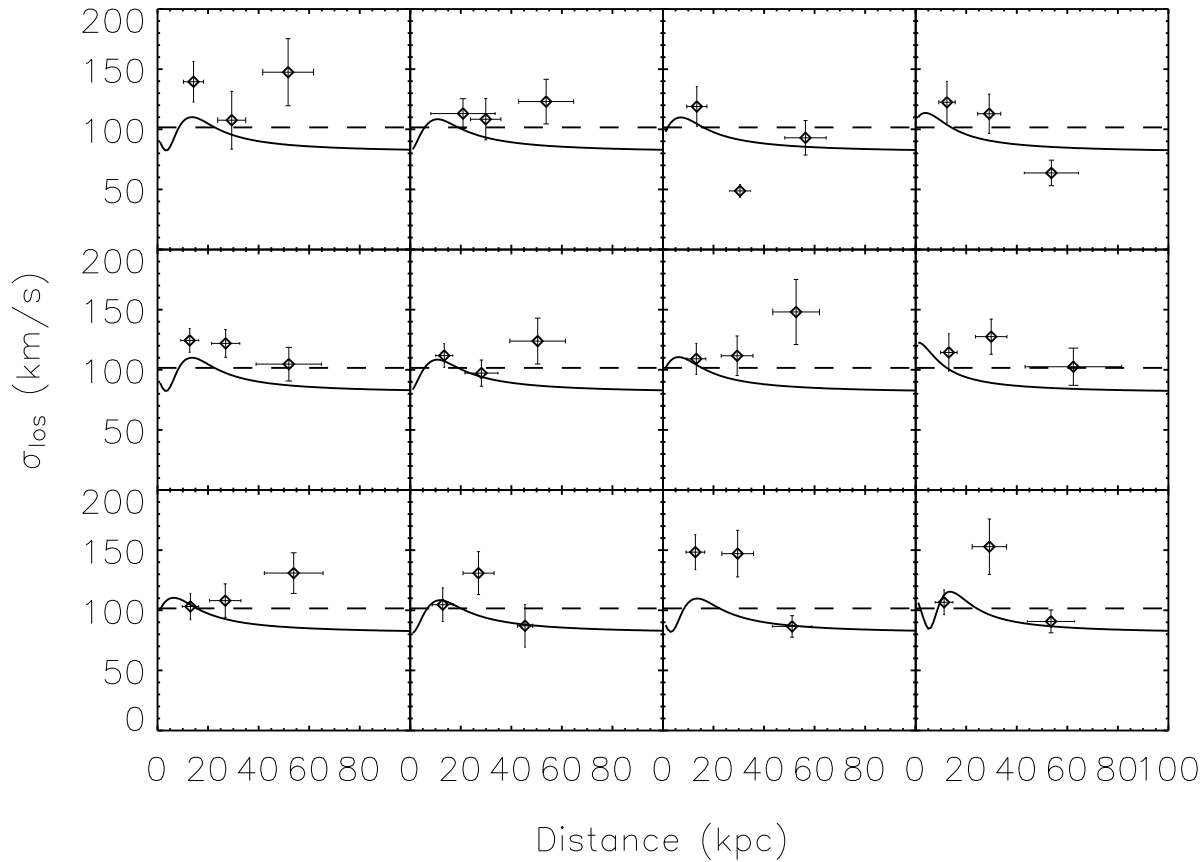,width=16cm}
    \caption{Comparison of \citet{sirko04a} data \siglos\ to the models fit
    by Sirko \etal~(dashed lines) and SLFC (solid lines). Velocities are
    heliocentric and distances are in kpc.}
    \label{fig: sdss-sigma-plots}
  \end{center}
\end{figure*}

We analysed the kinematics of our stars following the methodology of
SLFC. This involved dividing the sample into regions on the sky in which
there are sufficient stars to measure \siglos\ for two distance bins along
the line of sight. The stars in each field were split evenly into the two
distance bins. The requirement we applied, that each bin should have at
least twenty stars, forced us to exclude all stars with Galactic latitude
$b < 0\degree$. This left us with a total of \num~stars with which to study
the kinematics of the Galactic halo. The distribution on the sky of all
\num~FHB stars in our final sample is shown in Fig.~\ref{fig: allsky-lb}.
The average distance to the stars is 6.7\kpc; the distance distribution is
shown in Fig~\ref{fig: dist-hist}.

In calculating the velocity dispersion along the line-of-sight (\siglos) we
excluded the upper and lower 5\,\% the data, in order to protect these
numbers from the adverse effects of outliers.  The result is adjusted by a
factor of $\frac{1}{0.789}$ to recover the true standard deviation of the
original distribution, a process described in detail by
\citet{morrison90-datatrim}.  The \siglos\ were corrected for a
30\kms~velocity measurement error by subtraction in quadrature and then
compared to the model predictions.  We have consistently used velocities in
a heliocentric frame throughout the calculations, with the corresponding
distance. Since the model predictions are made in a galactocentric frame, a
projection factor was applied to recover the appropriate line-of-sight
value.

%%%%%%%%%%%%%%%%%%%%%%%%%%%%%%%%%%%%%%%%%%%%%%%%%%%%%%%%%%%%%%%%%%%%%%%%

\section{Results and Analysis}
\label{sec: results}

Table~\ref{tab: fields-sigma}, in addition to listing the fields, shows the
observed velocity dispersions and the 1-$\sigma$ errors.  These have been
used to test both an isothermal, isotropic halo with velocity dispersion of
101.6\kms~(i.e., the value observed by \citet{sirko04b}) and the
predictions of the SLFC model.  Fig~\ref{fig: sigma-plots} shows this
comparison for all six northern fields.  The SLFC predictions are shown as
the solid line, while the isothermal model is the horizontal dashed
line. Visual inspection leads us to conclude that both models fit the data
equally well; a \chisq~analysis confirms this view.  Both models arise from
predictions rather than fits to the data, and hence have no free
parameters. For the twelve \siglos\ in the six fields we measured a reduced
\chisq~of 1.31 and 1.85 for the SLFC and isothermal models respectively,
with 12 degrees of freedom each.  Since this is not a statistically
significant difference, both must be regarded as fitting the data equally
well.

\begin{table}
  \caption{Definition of fields in Galactic co-ordinates, showing Bottom Left
  Corner (BLC), Top Right Corner (TRC), number of stars, distance and
  \siglos\ of stars in field. Errors are 1$\sigma$.}
  \label{tab: fields-sigma} 
  \begin{center}
    \begin{tabular}{|l|c|c|c|r|c|}
%  \leavevmode
    \hline
	\hline
    Field   & BLC        & TRC        &  Num&     Dist        & \siglos \\
            & $(l,b)$    & $(l,b)$    &     &     (kpc)       & \kms  \\
    \hline
    Field 1 & (220,45)   & (275,70)   & 22  & $ 6.3 \pm 1.2$  & $ 108 \pm 17$ \\
            &            &            & 22  & $11.5 \pm 2.3$  & $ 110 \pm 17$ \\
    Field 2 & (275,45)   & (320,70)   & 78  & $ 5.2 \pm 1.0$  & $ 115 \pm 10$ \\
            &            &            & 79  & $ 9.0 \pm 1.8$  & $ 125 \pm 10$ \\
    Field 3 & (320,45)   & (360,70)   & 31  & $ 5.5 \pm 0.9$  & $  92 \pm 12$ \\
            &            &            & 32  & $ 8.9 \pm 1.7$  & $ 120 \pm 15$ \\
    Field 4 & (220,20)   & (275,45)   & 24  & $ 5.0 \pm 1.2$  & $ 104 \pm 16$ \\
            &            &            & 24  & $ 8.7 \pm 1.0$  & $ 149 \pm 22$ \\
    Field 5 & (275,20)   & (320,45)   & 44  & $ 3.9 \pm 0.8$  & $ 113 \pm 13$ \\
            &            &            & 45  & $ 7.5 \pm 1.8$  & $ 120 \pm 13$ \\
    Field 6 & (320,20)   & (360,45)   & 64  & $ 4.3 \pm 0.8$  & $  84 \pm  8$ \\
            &            &            & 65  & $ 8.2 \pm 1.9$  & $  98 \pm  9$ \\
    \hline
  \end{tabular}
\end{center}
\end{table}

\subsection{Outer Halo Comparison}

\citet{sirko04a} have provided a similar sample of 1170 FHB stars in the
outer halo (distances up to $\sim 100\kpc$). From this sample they derive
an estimate of the isothermal velocity dispersion of 101.6\kms.  We have
analysed their data in the same manner as above (see their Table 3 for
distances and heliocentric red-shift). The results are shown in
Fig~\ref{fig: sdss-sigma-plots}.  At the distance of the Sirko data, both
models predict similar line-of-sight velocity dispersions.  Therefore these
data are not an effective test of the SLFC model.  These data, divided into
nine fields and three distance bins per field, fit both models equally
well.  The reduced \chisq~statistics for the two models are 2.37 and 2.19
for the SLFC and isothermal models respectively, for 35 degrees of freedom,
excluding the obvious outlier in panel 3 (third from left, top row in
Fig~\ref{fig: sdss-sigma-plots}). This data point is likely caused by the
Sagittarius stream \citep{ibata94}, as noted by
\citeauthor{sirko04a}. Again, the difference in \chisq~is not statistically
significant.

\subsection{Discussion}
It is clear that neither this sample of inner Galactic halo stars, nor the
outer halo sample of \citet{sirko04a}, permit us to distinguish between the
two competing models.  It is worth noting here that the SLFC model is a
physically realisable system \citep{flynn96}, whereas the isothermal model
does not account for the locally observed halo velocity dispersion
anisotropy.  Where then should we look in order to best discriminate between
these two models?

\begin{figure}
  \begin{center}
    \leavevmode
    \epsfig{file=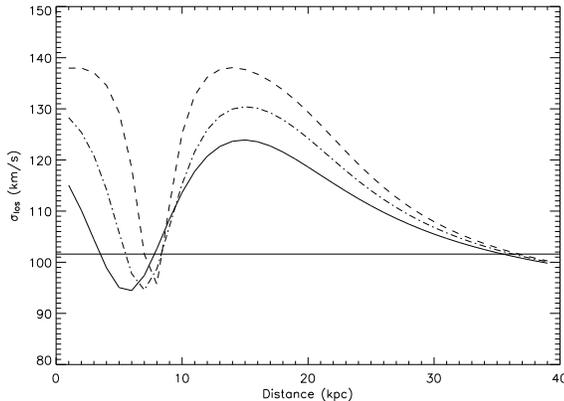,width=8cm}

    \caption{Difference between the SLFC and isothermal model predictions
    in the direction of the Galactic centre.  Dashed, dot-dashed and solid
    curves are drawn for lines-of-sight at $(l,b) = (0,15), (0,30), (0,45)$
    respectively.  For comparison, the solid line shows the constant
    velocity dispersion measured in the Sirko sample.}

    \label{fig: model}
  \end{center}
\end{figure}

We have used the model of SLFC to determine regions on the sky where the
two models differ in their predictions of \siglos\ by more than 30\kms.
This area is restricted to lines-of-sight toward the Galactic centre and at
low Galactic latitude.  Along these lines of sight one might expect a clear
signature of the SLFC model, as demonstrated in Fig~\ref{fig: model}.  A
survey of 500 FHB stars at $(l,b) =
(340\;\mbox{to}\;20;-30\;\mbox{to}\;+30)$ with distance bins centred at
$4$, $8$ and $13$\,\kpc~would provide a good discriminant between the
models.  This effect may also be seen in Field 6 of Fig~\ref{fig:
sigma-plots}, which is the closest the bulge.

%%%%%%%%%%%%%%%%%%%%%%%%%%%%%%%%%%%%%%%%%%%%%%%%%%%%%%%%%%%%%%%%%%%%%%%%

\section{Summary and Conclusions}
\label{sec: Conclusion}

Line of sight velocities have been measured for a sample of \num~FHB stars
in the inner Galactic halo, with an average heliocentric distance of
6.7\kpc. We have used these velocities to test the velocity dispersions
predicted by the \citet{slfc94} halo model, and compared with those
predicted by an isothermal model. The former model allows for marked
anisotropies to account for the well-constrained parameters of the local
Galactic halo. These new data are equally well described by both models. We
also compared the model predictions to the sample of 1170 FHB stars in the
outer Galactic halo, as published by \citet{sirko04a}.  We confirm their
findings that both models fit these data equally well.

We suggest that the best place on the sky to differentiate these two models
is towards the Galactic bulge and at low galactic latitudes; we propose
that future efforts concentrate on the more challenging task of recovering
clean samples of halo FHB stars at low galactic latitudes.

%%%%%%%%%%%%%%%%%%%%%%%%%%%%%%%%%%%%%%%%%%%%%%%%%%%%%%%%%%%%%%%%%%%%%%%%

\section*{Acknowledgements}
\label{sec: acknowledgements}

We thank the UKST staff for their support during the observations. We also
thank Lee Clewley and collaborators, who kindly provided the Sersic profile
fitting code.

CT would like to thank Tuorla Observatory for its support during an
extended stay, and Hamburger Sternwarte for support during several
visits. CF thanks the Academy of Finland and the ANTARES program for its
support of space based research. BKG, CF, and CT acknowledge the financial
support of the Australian Research Council and its Discovery and Linkage
International programs.
 
T.C.B. acknowledges partial support from grants AST 00-98508, AST 00-98549, and
AST 04-06784, as well as from grant PHY 02-16783; Physics Frontier Center/Joint
Institute for Nuclear Astrophysics (JINA), awarded by the U.S. National Science
Foundation.  We are also grateful to Michigan State University for providing
partial financial support for the UK Schmidt Telescope during the time
the data presented herein were gathered.

\bibliography{thom-references}
\bibliographystyle{mn2e}

\end{document}